\documentclass[manuscript,review=false,anonymous=false]{acmart}

\AtBeginDocument{%
  \providecommand\BibTeX{{%
    \normalfont B\kern-0.5em{\scshape i\kern-0.25em b}\kern-0.8em\TeX}}}

\setcopyright{acmcopyright}
\copyrightyear{2018}
\acmYear{2018}
\acmDOI{10.1145/1122445.1122456}

\usepackage{booktabs} 
\usepackage{url}

\usepackage{caption}
\usepackage[all]{nowidow}
\usepackage{wrapfig}
\usepackage{array}
\usepackage{arydshln}
\usepackage{tabularx}
\usepackage{multirow}
\usepackage{arydshln}
\newcolumntype{L}[1]{>{\raggedright\let\newline\\\arraybackslash\hspace{0pt}}m{#1}}
\newcolumntype{C}[1]{>{\centering\let\newline\\\arraybackslash\hspace{0pt}}m{#1}}
\newcolumntype{R}[1]{>{\raggedleft\let\newline\\\arraybackslash\hspace{0pt}}m{#1}}

\usepackage{wrapfig,lipsum,booktabs} 

\def\authnotes{1}
\newcounter{notectr}[section]
\newcommand{\thenote}{\thesubsection.\arabic{notectr}\refstepcounter{notectr}}

\newcommand{\note}[2]{$\ll$#1~\thenote: #2$\gg$}
\newcommand{\cnote}[1]{\ifnum\authnotes=1 \textcolor{blue}{\note{Comment:}{#1}}\fi}

\copyrightyear{2026}
\acmYear{2026}
\setcopyright{acmlicensed}\acmConference[CHI '26]{CHI}{Nov 5-9, 2026}{USA}
\acmBooktitle{CHI'24, Nov, 2026, USA}
\acmPrice{15.00}
\acmDOI{10.1145/3491102.43}
\acmISBN{978-1-4503-9157-3/22/04}

\begin{document}



\title[Data Protection]{When Data Protection Fails to Protect: Law, Power, and Postcolonial Governance in Bangladesh}


\author{Pratyasha Saha}
\affiliation{%
  \institution{University of Illinois Urbana Champaign}
  \country{USA}}
\email{saha19@illinois.edu}

\author{Anita Say Chan}
\affiliation{%
  \institution{University of Illinois Urbana Champaign}
  \country{USA}}
\email{achan@illinois.edu}

\author{Sharifa Sultana}
\affiliation{%
  \institution{University of Illinois Urbana-Champaign}
  \country{USA}}
\email{sharifas@illinois.edu}

\renewcommand{\shortauthors}{Saha et al.}

\begin{abstract}


Rapid digitization across government services, financial platforms, and telecommunications has intensified the collection and processing of large-scale personal data in Bangladesh. In response, the state has introduced multiple regulatory instruments, including the Personal Data Protection Ordinance, the Cyber Security Ordinance, and the National Data Governance Ordinance in 2025. While these initiatives signal an emerging legal regime for data protection, little scholarly work examines how these frameworks operate collectively in practice. This paper presents a legal and institutional analysis of Bangladesh’s emerging data protection regime through a systematic review of these three ordinances. Through this review, the paper provides an integrated mapping of Bangladesh’s evolving data protection framework and identifies key legal and institutional barriers that undermine the effective protection of citizens’ personal data. Our findings reveal that this emerging regime is constrained by limited institutional independence, uneven regulatory capacity, and the misaligned legal assumption of individualized, autonomous data subjects. Furthermore, these frameworks invisibilize prevalent sociotechnical layers, such as informal data flows and mediated access via “human bridges”, rendering formal protections difficult to operationalize. This paper contributes to HCI scholarship by expanding the concept of data protection as a complex sociotechnical design problem shaped by the informal infrastructures of the Global South.

\end{abstract}








\settopmatter{printfolios=true}

\maketitle

\section{Introduction}

Over the past decade, Bangladesh has rapidly expanded digital governance initiatives through programs such as Digital Bangladesh~\cite{DigitalBangladesh}, large-scale biometric identity systems~\cite{Macdonald_2021}, and the growth of mobile financial services used by tens of millions of citizens~\cite{mfsgrowth}. These infrastructures process vast amounts of personal data, including demographic information, biometric identifiers, financial records, and communication metadata. As the digital ecosystem grows, concerns surrounding privacy, surveillance, and data misuse have increasingly drawn attention from scholars, policymakers, and civil society organizations~\cite{bdsurveillance, 165ahmed2017privacy}. In July 2023, a vulnerability in Bangladesh government's Birth and Death Registration system exposed the personal information of over 50 million citizens, including names, phone numbers, and national identity numbers~\cite{Franceschi-Bicchierai_2023}. This database could be accessed through publicly available APIs, which revealed sensitive demographic information at national scale. More recently, a technical failure in the Election Commission’s website exposed the personal data of around 14,000 journalists, including national ID numbers and mobile phone details, which demonstrates persistent vulnerabilities in government data management systems~\cite{journadata}. 

Beyond data breaches, concerns have also been raised about the broader governance environment surrounding digital regulation in Bangladesh. Investigative reports have revealed that some government employees illegally accessed and sold citizens’ personal data including NID records and phone call details, through hundreds of social media groups and online platforms~\cite{datasell}. These incidents demonstrate how personal data frequently circulates through informal and mediated networks that remain largely invisible within formal regulatory frameworks such as ICT Act~\cite{ict}, Telecommunications Regulation Act, Personal Data Protection Ordinance~\cite{bdpdpo}, Cyber Security Ordinance~\cite{bdcso}, and National Data Governance Ordinance~\cite{bdndgo}. These laws were recently amended by the Bangladesh government to regulate the collection, processing, and sharing of personal data, yet scholars and activists warn that these emerging legislations fail to incorporate adequate safeguards and oversight mechanisms to ensure robust privacy protections for citizens~\cite{eusee2025, techglobal, hrwcrit, deshkal, tib}.

To understand this mismatch between policy frameworks and the ways data actually circulates in practice, we draw on Arora’s critique of aspirational digital governance. Arora argues that digital governance is often framed through narratives of technological modernization, where regulatory frameworks assume forms of digital participation that mirror idealized technological futures rather than the sociotechnical realities through which digital systems actually operate~\cite{aroraprivacy}. In this framing, governance systems are designed around imagined infrastructures of orderly data flows, institutional capacity, and individualized participation, even when everyday digital practices are mediated through informal networks, infrastructural improvisation, and relational forms of access.

Within HCI and ICTD scholarship, however, how data protection regimes operate as sociotechnical systems remains understudied. Existing research often focuses on normative privacy principles or high-level regulatory reforms without examining how multiple laws interact or what kinds of institutional and sociotechnical infrastructures they assume for enforcement~\cite{nouwens2020dark, utz2019informed, machuletz2019multiple}. As a result, data protection regimes are frequently treated as formal legal frameworks rather than as governance systems embedded within broader institutional and infrastructural contexts~\cite{wong2019bringing}. Our research addresses this gap through a qualitative legal and institutional analysis of Bangladesh’s emerging data protection regime. We examine three key regulatory instruments: the Personal Data Protection Ordinance~\cite{bdpdpo}, the Cyber Security Ordinance~\cite{bdcso}, and the National Data Governance Ordinance~\cite{bdndgo}, to understand how they conceptualize personal data governance and what institutional or sociotechnical infrastructures they assume for enforcement. Our broader goal is to understand why, despite the growing presence of formal legal protections, citizens’ personal data remains vulnerable in practice. Specifically, we address three research questions:
\begin{itemize}    
     \item \textbf{RQ1: What is the legal regime of data protection in Bangladesh?}
     \item \textbf{RQ2: What forms of institutional and technical infrastructure do these frameworks assume for enforcement?}
     \item \textbf{RQ3: Why do existing data protection laws in Bangladesh fail to function as effective protections for citizens in practice?}
\end{itemize}

Our analysis reveals several structural weaknesses within the current legal framework. Although multiple ordinances reference principles of data protection and digital governance, they distribute regulatory authority across overlapping agencies while granting broad discretionary powers to state actors. Institutional oversight mechanisms remain weak or ambiguously defined, and the frameworks assume enforcement capacities, such as independent regulatory bodies, technical compliance monitoring systems, and transparent accountability mechanisms, that are largely absent in practice. Most significantly, the framework renders invisible the critical sociotechnical layers of data handling in citizens’ interactions with digital systems, including informal intermediary bodies that remain unrecognized within the law’s regulatory categories - actors we define as ``human bridges". Thus, the legal regime constructs a model of data governance that appears comprehensive on paper yet fails to account for both the institutional limitations and the sociotechnical practices through which personal data is accessed, processed, and shared in practice.

Thus, this paper makes a threefold contribution to the HCI and ICTD literature. First, we map the emerging data protection regime in Bangladesh by examining how multiple legal instruments collectively structure personal data governance and what institutional capacities they assume for enforcement. Second, we show how personal data governance in practice is mediated through sociotechnical infrastructures like ``human bridges" that remain largely invisible within formal regulatory frameworks. Third, we advance the decolonial HCI scholarship by showing how universalized regulatory models can obscure the relational infrastructures through which digital participation and data governance operate in reality, and thus limit the effectiveness of formal regulatory protections.

\section{Background}

\subsection{Privacy, Data Protection and AI Systems}

A growing body of work in privacy, law, and HCI has examined the implications of proliferating data collection and AI-driven systems for personal data protection. Advances in large-scale data aggregation, algorithmic inference, and computational power have made it increasingly feasible to reconstruct sensitive information about individuals from publicly available or seemingly innocuous data~\cite{barocas2014big, rocher2019estimating, lermen2026large}. These developments have undermined the effectiveness of conventional privacy-preserving approaches such as anonymization, de-identification and aggregation, and raised intensified concerns around data misuse, surveillance, and loss of individual autonomy~\cite{cohen2012configuring, zuboff2023age, susser2019technology}. 

On the technical front, this body of work calls for more robust privacy-enhancing technologies (PETs), explicit accounting of cumulative privacy loss, and rigorous transparency in data processing practices. A significant line of research has advanced formal privacy guarantees, most notably differential privacy (DP), which mathematically bounds the risk of individual re-identification~\cite{dwork2014algorithmic, abadi2016deep}. However, recent scholarship highlights how DP introduces severe operational friction and systemic usability challenges in translating theoretical privacy assumtions into practical system configurations~\cite{haeberlen2011differential, gurses2016privacy, ngong2024evaluating}. In regard to minimizing centralized data collection, researchers have introduced decentralized paradigms like federated learning which train models locally~\cite{kairouz2021advances}. Yet, recent security evaluations demonstrate that such decentralized approaches remain highly vulnerable to sophisticated reconstruction and membership inference attacks~\cite{nguyen2022flame, mothukuri2021survey}. Concurrently, the emerging field of machine unlearning attempts to operationalize the legal `right to be forgotten' within generative models~\cite{bourtoule2021machine, eldan2024s, jang2023knowledge, gandikota2023erasing}. State-of-the-art research in this domain cautions that `inexact unlearning' methods often fail to completely excise user data, necessitating new adversarial frameworks to rigorously measure sample-level unlearning completeness~\cite{cooper2024machine}. As the limitations of purely technical fixes become apparent in this ongoing arms race of building robust data protection infrastructures, systems and HCI research have evolved to incorporate socio-technical governance frameworks into the design and deployment of real-world data pipelines.

Technical data protection and privacy mechanisms rely on simplifying assumptions about adversaries, data distributions, and system boundaries that often do not hold in practice~\cite{narayanan2010myths, bambauer2013fool, kifer2011no, rakova2021responsible}. In HCI, scholars have argued that privacy risks do not arise solely from data release, but emerge through downstream uses, institutional practices, and entrenched power asymmetries that cannot be mitigated through technical controls alone ~\cite{groverprivacy, solove2010understanding, nissenbaum2009privacy, cohen2019turning}. Critical HCI scholarship conceptualizes privacy not as a static property of datasets, but as a situated and relational practice shaped by social norms, infrastructures, and governance regimes~\cite{dourish2018datafication, boydnetwork, aroraprivacy}. This approach positions data protection as a central regulatory response to mitigate the risks associated with data collection, use, and dissemination~\cite{hoofnagle2019european}

\subsection{Data Protection and Governance}

Data protection has become a central pillar of contemporary governance, with the European Union’s General Data Protection Regulation (GDPR) widely recognized as the most influential model shaping global regulatory standards~\cite{Bradford_2020}. Since its enforcement in 2018, GDPR, alongside frameworks such as the California Consumer Privacy Act (CCPA), has driven the worldwide diffusion of data protection principles~\cite{Greenleaf_2021}. States and organizations have adopted GDPR-like approaches to regulate the collection, processing, and circulation of personal data~\cite{Greenleaf_2021}. Beyond formal law, this diffusion extends through corporate compliance frameworks that standardize how personal data is governed across contexts~\cite{Waldman_2021}.

Nonetheless, a growing body of work highlights that these governance models generate significant tensions in practice. Expanding digital governance agendas often produce overlapping legal regimes and interpretive ambiguities, requiring additional meta-regulatory or co-regulatory mechanisms to resolve. For example, studies of GDPR’s “legitimate interest” basis, show how firms exploit legal flexibility through deceptive interface design, enabling extensive personal data processing while maintaining formal compliance~\cite{Kyi_AmmanaghattaShivakumar_Santos_Roesner_Zufall_Biega_2023, Nouwens_Liccardi_Veale_Karger_Kagal_2020}.

Comparative research further demonstrates that the global adoption of GDPR-like frameworks does not ensure effective or equivalent protection~\cite{Taylor_2017}. Studies across regions such as Latin America, Africa, and Asia show that transplanted legal models are reinterpreted through local political economies, institutional capacities, and cultural understandings of data, ownership, and state authority~\cite{CarrilloJackson, GrisalesRendn_2022, greenleaf2014asian, Ekpo_Okokon_Akpakpan_2024, Milan_Trer_2019, Okolo_Aruleba_Obaido_2023}. Empirical work on emerging AI ecosystems such as in the Global South, rapid adoption of AI systems coexists with limited capacity and uneven technical literacy, which expose a fundamental mismatch between globally circulating governance models and locally situated realities~\cite{sultana2026}.

These limitations can be understood through broader theories of governance and infrastructure. Modern regulatory systems rely on simplifying complex social realities into standardized, legible categories such as “data subject,” “data controller,” and “consent” that assume orderly, transparent, and institutionally bounded data flows~\cite{scott2020seeing}. In practice, however, data ecosystems are far more fragmented, opaque, and socially embedded than these categories account for. In broader arguments in HCI and decolonial computing, Dourish and Mainwaring~\cite{118dourish2012ubicomp} describe this universalization of technical and governance models as a “colonial impulse,” where standardized frameworks are projected onto diverse sociotechnical contexts without accounting for local infrastructures or practices. 

While this body of work established that globally circulating data protection models often fail to translate effectively across contexts, it offers limited insight into \textit{why} these failures occur in practice. Our study explicitly engages this “why” by outlining how the design elements of data protection frameworks overlook fundamental institutional, social and relational structures, which function as critical pivot points where formal regulatory assumptions encounter lived realities and begin to fail.

\section{Methodology}

The goal of this study was to examine how Bangladesh’s emerging data governance framework constructs and operationalizes personal data protection in practice. For this, we (1) analyzed the legal structure of the country’s current data protection regime, (2) examined the institutional and technical infrastructures assumed by these laws for enforcement, and (3) identified structural conditions under which these legal protections fail to function as effective safeguards for citizens. To fulfil this goal, we conducted a qualitative legal–institutional analysis of three recently introduced regulatory instruments in Bangladesh: the Personal Data Protection Ordinance (2025), the Cyber Security Ordinance (2025), and the National Data Governance Ordinance (2025).

\subsection{Data Collection}

We started by identifying and collecting primary legislative texts that currently define the country’s emerging data governance framework. We collected the full statutory texts of the Personal Data Protection Ordinance (2025), the Cyber Security Ordinance (2025), and the National Data Governance Ordinance (2025). These instruments collectively establish the legal provisions governing personal data protection, cybersecurity oversight, and public-sector data governance in Bangladesh.

To contextualize the design and operational assumptions of these frameworks, we also reviewed adjacent legal and regulatory materials, including constitutional provisions related to data privacy, existing telecommunications and cybersecurity regulations such as Bangladesh Telecommunications Regulatory Act (BTRA) ~\cite{btra} and ICT Act~\cite{ict}, and publicly documented policy statements relevant to digital governance. All materials were collected from official government repositories, legal databases, and publicly accessible regulatory publications between November 2025 and January 2026.

Further, to situate the legal analysis within the broader governance landscape, we also reviewed publicly documented incidents related to personal data exposure, cybersecurity vulnerabilities, and data governance controversies in Bangladesh. These materials included investigative newspaper reports, publicly reported data breaches, regulatory announcements, and expert commentary discussing weaknesses in the protection of citizens’ personal information. The purpose of reviewing these materials was not to conduct a systematic incident analysis, but rather to understand the real-world contexts in which the examined legal frameworks are expected to operate. These contextual materials helped interpret how statutory provisions relate to practical governance challenges such as institutional capacity constraints, weak enforcement mechanisms, and recurring data security failures.

\subsection{Data Analysis}

We conducted an inductive thematic analysis~\cite{braun2012thematic} across data protection statutes, including full texts of enacted data protection laws, constitutional provisions related to data privacy, subsidiary regulations, and relevant adjacent legislation (e.g., ICT and telecommunication laws). 

For primary legal texts not available in English, we translated the text before analyzing. Personal Data Protection Ordinance (2025)~\cite{bdpdpo}, the National Data Governance Ordinance (2025)~\cite{bdndgo}, and Cyber Security Ordinance (2025)~\cite{bdcso} of Bangladesh were published only in Bengali. One of our authors, a native Bengali speaker, translated these documents into English in consultation with a local legal practitioner to ensure terminological accuracy and fidelity to legislative intent.

To guide the initial coding process, we started by applying an existing evaluative framework from recent FAccT scholarship~\cite{alanoca2025comparing} on comparative AI regulation governance. Initially we started mapping statutory provisions to the framework’s core dimensions, including regulatory scope, rights articulation, institutional design, enforcement mechanisms, and accountability structures. However, during the coding process we observed recurring patterns that were not fully captured by these predefined categories. In particular, provisions related to state exemptions, executive override powers, institutional capacity constraints, and sociotechnical barriers to exercising legal rights appeared repeatedly across the analyzed statutes. Instead of treating these patterns as anomalies, we treated the base framework as an initial diagnostic scaffold and reorganized our analysis inductively around recurring structural themes observed in the legal texts. We followed open coding~\cite{75strauss1990open} for analyzing the statutory provisions and iteratively clustered the codes into higher-level analytic themes. During the process, we had regular discussions with the research team members to collectively develop the codebook and finalize the themes. 

\subsection{Development of Structural Failure Logics}

Through this iterative analysis, we identified several recurring structural dynamics that shape how data protection laws are assumed to operate in practice. We conceptualize these dynamics as structural failure logics: patterns within legal design that limit the ability of formal rights protections to function as meaningful constraints on data collection, processing, and state access. These logics are not intended to capture every feature of the statutes. Instead, they isolate structural conditions under which data protection regimes risk becoming symbolic protections rather than operational safeguards.

The analysis ultimately converged around three interrelated axes: \begin{itemize}
    \item \textbf{Authority, Discretion, and the Fragility of Rights}: The first axis examines authority, discretion, and the fragility of rights. This dimension analyzes whether the legal framework meaningfully constrains state access to personal data or instead formalizes discretionary authority through broad exemptions, override provisions, and security carve-outs.
    \item \textbf{Institutional and Infrastructural Capacity}: The second axis focuses on institutional and infrastructural capacity. This dimension assesses whether statutory obligations assume regulatory, judicial, and technical capacities that plausibly exist in practice, including independent regulatory bodies, enforcement personnel, digital forensic infrastructure, and adjudicatory mechanisms.
    \item \textbf{Rights Practicality and Social Access}: The third axis examines rights practicality and social access. This dimension evaluates whether legal rights such as consent, access, correction, and redress are realistically exercisable within the socio-economic and institutional conditions of Bangladesh.

These axes function as analytic lenses that allow us to examine how Bangladesh’s emerging data protection regime operates within broader governance ecosystems and why formal legal protections fall short to translate into meaningful protection for citizens in practice.

\end{itemize} We present these axes as analytic lenses rather than evaluative checklists. They enable a systematic examination of how data protection laws operate within broader governance ecosystems and why formal legal protections may fail to translate into lived protection.

\begin{figure}
    \centering
    \includegraphics[width=1\linewidth]{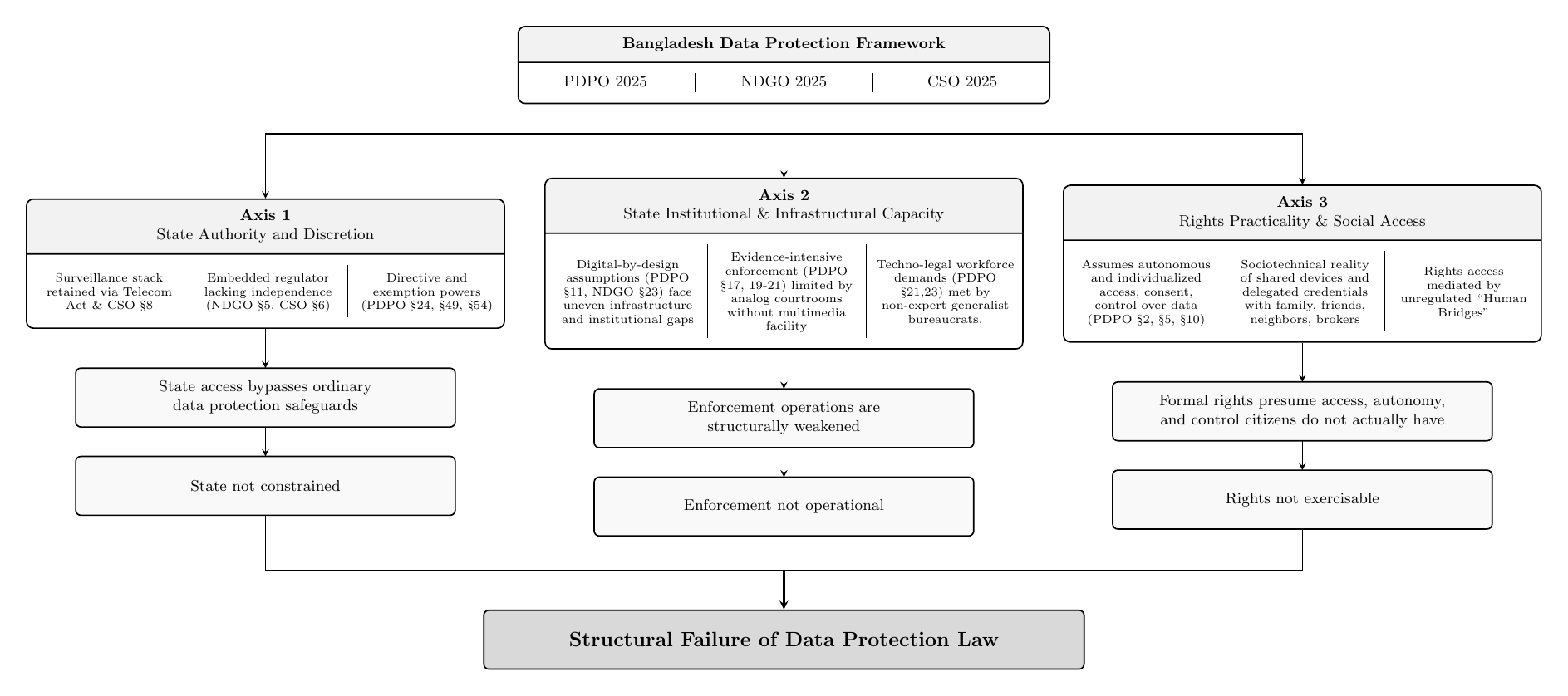}
    \caption{Figure 1 models the structural failure logics of Bangladesh’s data protection regime. Rather than a single point of breakdown, failure emerges from the interaction of three systemic conditions: (1) unconstrained state authority that enables parallel access pathways, (2) weak institutional and infrastructural capacity that renders enforcement non-operational, and (3) a fundamental mismatch between individualized rights in law and socially mediated data practices in reality. Together, these conditions produce a regime in which data protection remains largely symbolic rather than a meaningful operational safeguard.}
    \label{fig:structural_failure}
\end{figure}

\section{Findings}

\subsection{Authority, Discretion, and the Fragility of Rights}

\subsubsection{Regulator Independence as a Structural Constraint}

In Bangladesh, the emerging data protection and cybersecurity regime adopts rights-oriented language while structurally embedding its enforcement institutions within executive authority. The legal framework positions regulatory bodies in ways that constrain their capacity to operate as independent oversight mechanisms over state data processing. 

The Personal Data Protection Ordinance, 2025 (PDPO)~\cite{bdpdpo} does not establish a standalone privacy regulator. Instead, it sets up the National Data Governance Authority as a statutory body “attached to the Prime Minister/Chief Advisor’s Office” (PDPO §8(2)). This arrangement places the institutional foundation of data protection governance within a broader executive-led data governance structure rather than creating an autonomous regulatory body dedicated to privacy protection. At the center of this structure is the National Data Governance Policy Board, which is chaired by the Prime Minister or Chief Advisor (NDGO §5(1). The composition of the Board is dominated by executive officials, including ministers, the Cabinet Secretary, and the Principal Secretary to the Prime Minister or Chief Advisor (NDGO §5(1)). Within this structure, representation from outside the executive is minimal: the law provides for only one representative nominated by the Government from civil society or human rights-related organizations (NDGO §5(1)(r)). This position also lacks institutional insulation, as the Government retains the authority to remove this member prior to the completion of their tenure (NDGO §5(3)). 

Under the CSO, the statutory entity for monitoring telecommunications data will be the Centre for Information Support (CIS) - an intelligence-gathering arm of the state responsible for internal security and law enforcement (Section 97). This institution would be established within the home ministry, and will have a three-member "quasi-judicial council" tasked with reviewing, approving, amending, or rejecting interception requests. However, the council is to be chaired by the minister of law, justice and parliamentary affairs, and will include the principal staff officer as well as the secretary of the home ministry as members. This structure also does not represent an independent oversight authority beyond state interference.

Beyond board composition, the legal framework provides the executive with direct authority over regulatory decision-making. Under the NDGO, the Prime Minister or Chief Advisor may issue immediate decisions and directions in emergency situations concerning matters governed by the ordinance (NDGO §7). Similarly, the Cyber Security Ordinance, 2025 establishes the National Cyber Security Agency as an “attached department” under the Information and Communication Technology Ministry (CSO §5(3)). Senior officials of the agency, including the Director General and Directors, are appointed by the Government and operate as full-time public officials executing functions assigned by the State (CSO §6(2)).

\subsubsection{State Exemptions and Executive Overrides}

The data protection framework incorporates extensive state exemptions and executive directive powers that significantly shape how the law operates in practice. While these provisions are justified in terms of national security, public order, and governmental administration, they create legal pathways through which state data processing may occur outside the ordinary regulatory safeguards established for private actors. Although the legislation references principles such as necessity and proportionality, it does not define or establish procedural safeguards such as independent judicial authorization, warrant requirements, mandatory logging of access, or notification to affected data subjects.

The Personal Data Protection Ordinance 2025 (PDPO) authorizes binding government directions to the regulatory authority responsible for implementing the law. Under Section 49, the Government may issue directives on grounds including sovereignty, national security, friendly relations with foreign states, and public order. The Authority is legally obligated to comply with these directives (PDPO §49(2)). In addition, Section 54 allows the executive to issue emergency orders related to data processing, enabling immediate intervention by the Government without clearly defined statutory thresholds or explicit provisions for judicial review.

Beyond directive powers, the PDPO also incorporates substantive exemptions for state data processing. Section 24 excludes several categories of governmental activity including national security, public order, law enforcement, taxation, and other administrative functions from the law’s core obligations governing consent and lawful processing. These provisions effectively remove large segments of state data processing from the regulatory scope applied to other actors. The ordinance also allows for additional exemptions to be defined through subordinate legislation, granting the executive branch considerable flexibility in determining when data protection obligations apply.

\subsubsection{Surveillance Stack Interaction Risk}

Bangladesh's data protection framework operates alongside a set of pre-existing surveillance, telecommunications, and criminal procedure laws that continue to govern the interception, monitoring, and disclosure of communications data. These sectoral regimes retain operational authority over the infrastructure through which state surveillance is conducted. As a result, many forms of state access to personal data remain regulated primarily through telecommunications and cybersecurity legislation rather than through the data protection framework established by the PDPO. This interaction is particularly visible within the country’s telecommunications and surveillance regime. Telecommunications and cybersecurity laws authorize interception and monitoring of communications networks, including access to traffic data, communications content, and network infrastructure, on grounds such as national security, public order, and public safety (Bangladesh Telecommunication Regulation Act Section 97). These powers are embedded within regulatory frameworks that historically govern telecommunications operators and network infrastructure. Because these authorities operate through sectoral legislation that predates the PDPO, the practical regulation of state surveillance continues to occur largely through these existing mechanisms rather than through the safeguards established within the data protection framework.

The Cyber Security Ordinance 2025 further expands the legal basis for state access to digital data. The ordinance permits interception of, or access to, traffic data when authorities have “reason to believe” that an offense has occurred, is occurring, or may occur (CSO Section 35). This evidentiary threshold enables investigative and preventative surveillance activities that operate outside the consent-based and rights-oriented mechanisms established by the PDPO. Consequently, state access to communications and network data can be authorized under cybersecurity and telecommunications legislation without triggering the procedural protections associated with personal data processing under the data protection framework. 

The coexistence of these legal regimes creates a layered governance structure in which the data protection law functions primarily as a framework for regulating civilian and commercial data processing, while state surveillance practices continue to be governed through sectoral security legislation. This institutional arrangement limits the extent to which the PDPO alone can operate as a comprehensive constraint on government access to personal data.

\subsection{Infrastructural and Institutional Limitations}

\subsubsection{“Digital-by-design” rights, analog-by-reality institutions}

Bangladesh’s emerging data protection framework is built around a technologically sophisticated model of rights enforcement that assumes the existence of interoperable national data infrastructure, reliable digital records, and technically capable regulatory institutions. In practice, however, the institutions responsible for implementing and enforcing these rights operate within administrative and technical environments where such infrastructure remains uneven. As a result, many of the rights formally guaranteed by the legal framework depend on institutional and technological capacities that are not yet fully established.

The Personal Data Protection Ordinance 2025 (PDPO) envisions a highly interconnected data governance architecture. For example, the law requires that data subjects be able to exercise rights such as access and portability through what the statute describes as a “Federated Interoperable Ecosystem” (PDPO §11(2)). Similarly, correction and erasure of personal data are designed to propagate across all relevant systems, including registries, backup systems, disaster recovery environments, and migration datasets (PDPO §14(6)). The law further requires that these changes be recorded within tamper-evident audit trails such as “immutable ledgers, blockchain, or equivalent technology” (PDPO §14(4)(b), §14(7)). These provisions assume the presence of sophisticated data synchronization systems and reliable cross-agency data governance infrastructure capable of maintaining consistent records across multiple technical environments.

The broader institutional framework outlined in the National Data Governance Ordinance 2025 similarly depends on advanced digital infrastructure. The ordinance proposes national-level systems such as the National Register of Data Exchange (NRDEX), described as a secure API-based layer enabling purpose-specific data sharing across agencies, as well as the Bangladesh National Data Integration Architecture (BNDIA), intended to coordinate data governance across government institutions (NDGO §23(d)). These systems form the technical backbone through which many of the rights envisioned under the PDPO would be operationalized.

However, the statutes themselves provide limited guidance on how these complex infrastructural systems will be developed, maintained, and operated in practice. The PDPO’s own definition of a data breach implicitly acknowledges potential infrastructural gaps, including breaches that occur due to “the absence of necessary equipment for proper processing and preservation” of data (PDPO §2(19)). This provision suggests an awareness that institutional and technological capacities required to implement the law may not yet be fully established.

A related concern lies in the institutional design of the regulatory bodies for data control and protection. PDPO, NDGO/CSO, and BTRA each establish their own regulatory bodies, respectively the National Data Governance Authority (NDGA), the Centre for Information Support (CIS), and the Bangladesh Telecommunication Regulatory Commission (BTRC). However, there is no clear specification of the legal structures or operational mechanisms through which these entities will coordinate and operate.

\subsubsection{Evidence-intensive enforcement without infrastructural capacity: “digital evidence admissible” is not “digital evidence usable”}

  Data protection enforcement is inherently evidence-intensive. Bangladesh’s data protection framework assumes that regulatory and judicial institutions can routinely collect, authenticate, and evaluate digital evidence generated by modern information systems. While the country’s legal framework has formally incorporated electronic records and computer-generated outputs into evidentiary practice, the practical capacity of enforcement institutions to process such evidence remains uneven~\cite{Uzzal_2025}.


 The evidentiary foundation for digital records in Bangladesh originates in the Information and Communication Technology Act, 2006, which recognizes electronic records and digital documents as legally valid forms of evidence~\cite{ict, evidencedigital}. More recent legislation has expanded the role of digital artefacts within investigative and enforcement processes. The Cyber Security Ordinance, 2025 authorizes authorities to collect and analyze digital logs, traffic data, and system records as part of cybercrime investigations~\cite{bdcso}. The Personal Data Protection Ordinance, 2025 (PDPO) further assumes that organizations will maintain extensive digital records demonstrating compliance with the law, including consent logs, processing records, breach notifications, and incident-response documentation (PDPO §§17, 19–21). These forms of evidence not only require technical infrastructure capable of securely preserving digital artefacts, also institutional expertise capable of interpreting them. However, the legal framework largely assumes the availability of such capabilities rather despite its absence.
 
 Evidence-intensive enforcement also presumes that investigative and judicial institutions possess the technical infrastructure required to present and evaluate digital artefacts in their original format. However, reports from legal practitioners, journalists, and court modernization initiatives indicate that many courtrooms in Bangladesh still lack basic multimedia infrastructure capable of displaying or examining digital evidence such as video recordings, CCTV footage, or platform-generated datasets.~\cite{Mogumder_2022, Uzzal_2025} Three things happen that hamper rights-enforcement: first, the evidence gets “flattened” into text descriptions or selective screenshots, which destroys context (timing, sequence, system context, interface behavior, and integrity metadata) and makes factual findings easier to manipulate; second, cross-examination becomes weaker because parties can’t point to the same artefact in real time; third, the process starts favoring litigants with resources; because the ability to repackage evidence into “court-friendly” formats (transcripts, certified extracts, curated exhibits) becomes a private capability rather than a public infrastructure guarantee~\cite{Uzzal_2025}. 

 These evidentiary limitations become particularly significant in the context of emerging synthetic media technologies. Deepfakes and other AI-generated manipulations allow audio, video, and images to be altered in ways that can closely mimic authentic recordings~\cite{Chesney_Citron_2019}. During the February 2026 election, Bangladesh experienced a significant proliferation of AI-generated misinformation, gendered abuse, and digitally mediated violence~\cite{Report_2026}. However, existing legal frameworks offer limited procedural guidance, technical standards, or institutional capacity to detect, attribute, and govern these forms of harm~\cite{aiwarn}. Bangladesh has not yet adopted a comprehensive regulatory framework addressing AI-generated media or establishing forensic protocols for authenticating synthetic digital content. In an environment where courts may already struggle to examine digital artefacts in their native formats, the growing availability of AI-generated media further complicates the evidentiary landscape.



 \subsubsection{The missing workforce}

 Bangladesh’s emerging data protection framework assumes the existence of a specialized techno-legal workforce capable of auditing complex digital systems, interpreting security architectures, and adjudicating evidence-intensive disputes involving data processing. However, the institutional structures responsible for implementing these laws remain largely staffed by generalist administrators and law-enforcement officials whose training is not oriented toward digital systems governance or forensic analysis.
 
The Personal Data Protection Ordinance, 2025 (PDPO) explicitly embeds these workforce assumptions within the legal framework. The ordinance mandates the creation of an audit panel composed of individuals with experience in information and communication technology, computer systems, personal data protection, or data privacy (PDPO §21(4)). The regulatory authority is further empowered to require data fiduciaries to undergo audits where data processing activities may pose risks to data subjects (PDPO §21(5)). In addition, organizations designated as “significant data fiduciaries” must appoint Chief Data Officers responsible for representing the organization before the Authority, facilitating the exercise of data subject rights, and ensuring timely remedies in cases involving misuse or mismanagement of sensitive personal data (PDPO §23(3)).

Beyond organizational roles, the ordinance presumes the existence of technical regulatory capacity within the Authority itself. The law authorizes the Authority to determine security standards for data processing (PDPO §17(4)) and to develop standard operating procedures covering key technical mechanisms such as consent management systems, data protection impact assessments, pseudonymization, erasure protocols, portability mechanisms, and cross-border data transfers (PDPO §27). Implementing and evaluating such measures requires regulators capable of interpreting security architectures, reviewing audit claims, and assessing system-level risk mitigation strategies.

In practice, however, Bangladesh’s data governance institutions have historically been staffed primarily by career bureaucrats, generalist administrators, and law-enforcement officials whose professional training is not centered on technology governance, policy, or law~\cite{Islam2024, Sabharwal_2013}. Investigative reports and legal analyses have repeatedly noted the limited availability of technical expertise within public institutions responsible for overseeing digital infrastructure and cybersecurity governance~\cite{Institute_2026}. As a result, many of the specialized roles assumed by the PDPO, including data protection auditors, technical regulators, and compliance officers, do not yet correspond to a clearly established professional workforce within the country’s governance ecosystem.

This institutional gap is reinforced by the structure of higher education and professional training in Bangladesh. Universities typically silo expertise into engineering, law, or public administration, with limited institutionalized pathways that combine these domains into an integrated field of technology governance or data protection law~\cite{sultana2026}. Consequently, enforcement institutions often draw from professionals trained in only one dimension of the problem: lawyers without technical systems knowledge, engineers without regulatory authority, and administrators without either.

\subsection{Rights Practicality and Social Access}

\subsubsection{Individualized Rights and the Social Reality of Digital Access} 
Data protection law in Bangladesh constructs rights around an individualized data subject capable of independently authoring consent, controlling digital accounts, and initiating legal remedies. This model assumes that personal data is generated, accessed, and managed by a single autonomous individual who exercises continuous control over devices, credentials, and communication channels. However, empirical research in ICTD and HCI has consistently shown that digital privacy practices in many contexts, including Bangladesh, is frequently organized around shared devices, delegated access, and relational forms of technology use rather than strictly individual control~\cite{Sambasivan2018, 129ahmed2017digital, sharingpasswordbrothers}.

The Personal Data Protection Ordinance 2025 (PDPO) grounds its rights architecture in this individualized conception of the data subject. The statute defines consent as a “clear, affirmative indication given by the data subject” (PDPO §2(22); §5), presuming that the person whose data is processed is also the individual directly interacting with digital systems and consent interfaces. Similarly, rights to access, correction, erasure, and objection must be exercised by the data subject through written requests directed to the data fiduciary (PDPO §10; §§31–35). Exercising these rights presumes that the data subject maintains continuous control over the device, account, or communication channel through which the request is made.

In practice, however, digital access in Bangladesh often involves shared or mediated use of technology. Mobile phones may be shared within households, SIM registrations may not correspond directly to the person using the device, and digital services are frequently accessed through intermediaries such as family members, local service agents, or community actors who assist with online interactions~\cite{cashlessbangladesh, digitalservice, patriarchy, computingstigma}. In such contexts, the person whose data is collected may not be the same person who controls the device, account credentials, or application interface through which data processing occurs. The PDPO does not explicitly recognize these forms of shared or mediated access. Its rights framework assumes stable authorship and control over digital identities, accounts, and communication channels. As a result, the legal mechanisms for exercising data protection rights may be difficult to operationalize in situations where digital access is distributed across multiple actors or mediated through social relationships.

\begin{figure}
    \centering
    \includegraphics[width=1\linewidth]{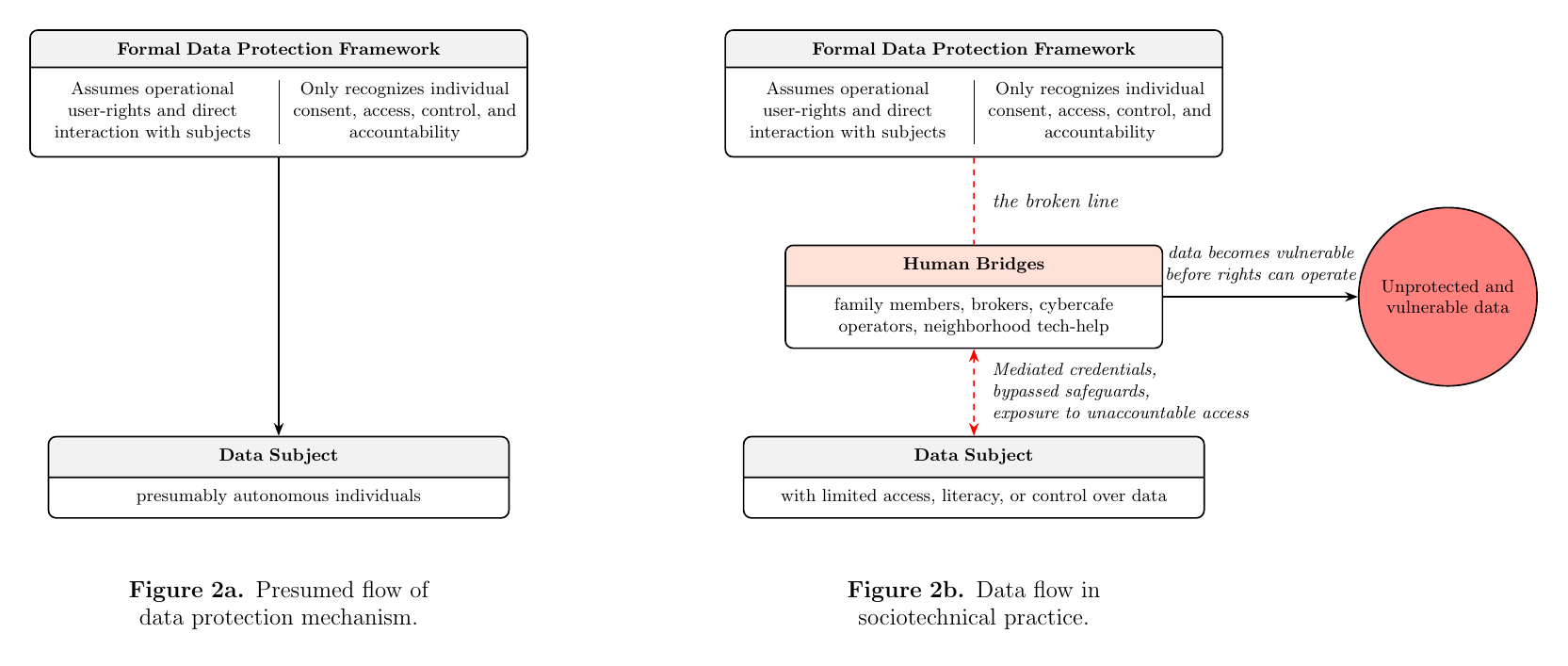}
    \caption{It illustrates the mismatch between formal data protection logics and their operational reality. Figure 2a shows the legal fiction of direct, rights-based interaction between an autonomous data subject and the data protection framework. Figure 2b demonstrates how this pathway breaks in practice: human intermediaries (``human bridges'') and mediated access infrastructures intervene, collapsing consent, weakening accountability, and exposing data to unregulated flows. Consequently, data becomes vulnerable prior to, and often outside the reach of, formal legal protections.}
    \label{fig:placeholder}
\end{figure}

\subsubsection{“Human Bridges” as Unregulated Data Actors} 

A cross-cutting limitation within Bangladesh’s data protection framework emerges from the presence of what this paper identifies as human bridges: informal intermediaries who make digital systems usable in practice but remain legally invisible within the regulatory architecture. These actors, family members, neighborhood shop owners, cybercafé operators, NGO workers, local brokers, and other “tech-help” figures, frequently assist individuals in navigating digital platforms, submitting online forms, uploading documents, and managing account credentials~\cite{digitalservice, cashlessbangladesh}. Prior ICTD research has shown that such intermediaries play an important role in enabling access to digital services in contexts where technological literacy, device ownership, or connectivity may be uneven~\cite{3sambasivan2010intermediated}. In effect, they form an informal data-processing layer that mediates interactions between individuals and digital systems.

However, Bangladesh’s Personal Data Protection Ordinance 2025 (PDPO) does not explicitly recognize these intermediated forms of digital access. The law assumes that interactions with digital services occur directly between a data subject and a data fiduciary. Consent is framed as an affirmative indication provided by the data subject, and the exercise of rights, such as access, correction, or erasure is similarly structured as a direct interaction between the individual and the organization processing their data. Outside narrow provisions related to guardianship for children or legally incapable persons (PDPO §9), the statute does not provide a framework for recognizing or regulating situations in which individuals rely on intermediaries to access digital systems.

In practice, however, many interactions with digital services occur through mediated access. Individuals may rely on trusted helpers to complete online applications, retrieve information from digital platforms, or manage accounts on their behalf. These intermediaries may temporarily store login credentials, upload personal documents, or interact with digital systems using accounts registered in another person’s name~\cite{digitalservice, computingstigma, 129ahmed2017digital}. While such practices often enable access to essential services, they also create situations in which personal data is processed or accessed by actors who fall outside the regulatory categories established by the law.

When harms arise through mediated access, such as misuse of credentials, coerced consent, or unauthorized disclosure of personal information, the allocation of responsibility becomes ambiguous. Because these intermediaries are invisibilized within the legal framework, the statute provides limited guidance on how accountability should be assigned when personal data is handled by individuals acting informally on behalf of others. As a result, responsibility may shift toward the data subject for failing to safeguard account credentials, even when access occurred through socially mediated assistance. Their legal invisibility within the data protection framework complicates questions of ownership, control, consent, and accountability.

\section{Discussion}

This paper examined how data protection regimes in Bangladesh adopt the language of rights while constructing enforcement, exception, and operational architectures that systematically undermine those rights in practice. The analysis revealed several structural tensions within the current framework: regulatory institutions remain closely integrated with executive authority, the enforcement model assumes technical and infrastructural capacities that are unevenly available, and the rights architecture overlooks the socially mediated nature of data practices around “human bridges” through which access, consent, and data flows are organized.

These findings suggest that the challenges surrounding data protection in Bangladesh arise from deeper mismatches between the legal model of data governance and the institutional, infrastructural, and social environments in which these laws operate. The framework articulates rights that depend on specialized regulatory expertise, robust digital infrastructures, and individualized patterns of technology use. Yet the institutions responsible for enforcing these rights operate within governance systems where techno-legal infrastructures are still evolving.



To interpret these patterns, we situate South Asian data protection laws within broader theoretical accounts of legibility, universalization, and decolonial computing~\cite{scott2020seeing, 118dourish2012ubicomp, aroraprivacy}. Read together, the three regimes function as legibility projects that translate privacy harms, data subjects, and compliance obligations into state-readable forms. These forms travel easily across jurisdictions and align with global data protection templates, but they do so by displacing the relational, mediated, and infrastructurally uneven conditions under which data access, harm, and contestation actually occur. This framing allows us to move beyond jurisdiction-specific doctrinal critique and thus argue for a situated governance framework aligned with sociotechnical realities.

\subsection{Rights as legibility projects}

Across the legal regime of data protection, rights are operationalized through procedures that privilege administrative tractability: standardized forms, identity verification requirements, documentary proof, and digitally preserved traces. These instruments are not inherently flawed; they reflect the state’s preferred techniques for making complex social phenomena governable. But this legibility is achieved by compressing social life into a narrow and idealized figure of the data subject: an individually authoring, continuously authenticated actor operating through stable interfaces and leaving behind clean evidentiary trails. This helps explain why “rights practicality” failures recur. 

The empirical finding we describe as “human bridges” identifies mediation as the dominant mode through which people access digital services and pursue remedies. Nevertheless, the legal regime treats these practices as structurally invisible within its rights machinery. In this regime, relational privacy is not encountered as a governance object in its own right, rather it is reorganized into forms that can be filed, verified, and adjudicated. In other words, these laws govern how privacy can be recognized as a legitimate claim - only through performances of identity, ownership, and procedure that the state can read. Scott’s account of legibility~\cite{scott2020seeing} clarifies the mechanism at work here. By translating privacy into standardized categories, these laws make data governance readable to the state while rendering relational and informal practices invisible or improper~\cite{111scott1998seeing}.


\subsection{Breach ecology and the collapse of rights under recurrent insecurity}


In Bangladesh, repeated large-scale data leaks and infrastructural failures reveal a broader pattern of systemic insecurity within the country’s digital governance ecosystem. These events illustrate what we describe as a `breach ecology': an environment in which recurring failures of data protection, weak institutional accountability, and limited avenues for redress interact to undermine the practical operation of privacy rights.

Reporting on data breaches involving biometric and identity-linked datasets illustrates how sensitive personal data circulates within an ecosystem where accountability mechanisms remain thin and avenues for individual redress are unclear~\cite{Khan_2024, Report_2026b, bddataleakwired}. These incidents demonstrate how high-value government and platform datasets may be exposed or redistributed through informal digital channels long before regulatory processes can respond. When breaches occur in environments with weak incident-response infrastructure, it becomes difficult to determine how data was accessed, which systems were affected, and who is responsible. If insecurity is persistent, the evidentiary foundations on which data protection enforcement depends are often the first elements to degrade or disappear. Without reliable evidentiary trails, individuals struggle to demonstrate harm, regulators struggle to attribute responsibility, and enforcement risks devolving into procedural compliance rather than meaningful accountability. In such environments, data protection law may continue to exist formally, but its ability to provide practical remedies becomes constrained by the very infrastructural fragilities it seeks to regulate.


\subsection{Implications for governance design}

Our findings suggest the need to move from rights-as-forms toward rights-as-relational governance. If mediated access is the dominant mode through which people interact with digital systems, then governance frameworks must explicitly account for mediation rather than treating it as fraud, exception, or invisibility. The “human bridges” identified in this study point to a regulatory target that is currently unaddressed: intermediaries who function as de facto processing layers but remain outside formal accountability structures.

The findings also highlight the importance of capacity sequencing. Interoperability-first imaginaries and digital-office adjudication models assume national-scale technical infrastructures and specialized workforces that cannot be taken for granted. Rights frameworks that depend on such infrastructures require explicit institutional preconditions; otherwise, the rights layer risks becoming a legitimacy veneer over uneven enforcement.

Finally, regulator independence emerges not as a normative aspiration but as a functional requirement in states that are themselves primary data processors. Where exemptions are broad and regulators are embedded within executive authority, data protection operates as an administrative program rather than a constraint on state power.

\subsection{Universalized data protection and Postcolonial HCI}

For HCI and ICTD, this paper contributes by explicitly connecting decades of empirical research on shared devices, mediated access, and infrastructural workarounds to the formal legal design of data protection rights. Prior work in HCI and ICTD has repeatedly demonstrated that digital access in many Global South contexts is relational, mediated, and deeply shaped by infrastructural scarcity and social interdependence~\cite{Sambasivan2019, toyama2011technology,burrell2016machine, 129ahmed2017digital, 3sambasivan2010intermediated}. This study adds an account of how such failures are structurally produced when governance frameworks are designed around an imagined user and institutional environment that does not exist. When policy assumes an individualized, continuously authenticated data subject and a high-capacity enforcement apparatus, it forecloses the very practices through which people actually exercise agency. In such settings, neither interaction design nor legal remedy can function as intended because both are anchored to a fictional governance subject.

More broadly, this paper situates Bangladesh’s data protection regime within a longer history of legal transplantation and regulatory mimicry. Policymakers have drawn heavily on GDPR-style rights frameworks (individual consent, purpose limitation, breach notification, and auditability) as markers of regulatory modernity and global alignment~\cite{Greenleaf_2021}. This pattern reflects what Dourish and Mainwaring describe as the colonial impulse of technolegal governance: the assumption that universalized models of order can be applied across diverse social and infrastructural contexts without attending to local practices of mediation, coordination, and repair~\cite{118dourish2012ubicomp}. Comparative scholars describe this pattern as isomorphic adoption, where legal forms are borrowed to signal legitimacy rather than to resolve locally situated governance problems. Payal Arora’s critique of “future-facing” digital rights imaginaries further clarifies this dynamic~\cite{aroraprivacy}. As Arora argues, governance in the Global South is frequently framed as a project of catching up to a technologically advanced future, where Western legal and technological frameworks are positioned as aspirational endpoints rather than contextually grounded interventions~\cite{arora2019general}. In this narrative, infrastructural unevenness is treated as a temporary deficiency that will disappear once societies become sufficiently digitized.

Our findings also demonstrate the consequences of this orientation. By universalizing an individualized rights-holder while assuming a high-capacity regulatory state, Bangladesh’s data protection framework renders mediation largely invisible while simultaneously making accountability depend on infrastructures such as secure digital records, reliable authentication systems, independent oversight mechanisms that remain uneven and fragile in practice. When these infrastructures fail or degrade, the evidentiary foundations required to exercise rights or pursue remedies may disappear with them.

For HCI and decolonial computing, the implication is not simply that policy should become “more context-aware.” Rather, our findings show that data protection frameworks systematically invisibilizes the sociotechnical layers through which data is actually accessed, shared, and governed in practice. In contexts such as Bangladesh, these omissions are compounded by limited institutional capacity, uneven technical infrastructures, and resource constraints that shape how data governance can be enacted. This suggests that rather than treating mediation, infrastructural limits, and institutional constraints as downstream implementation challenges, they must be understood as first-order conditions for designing data protection regimes. Without this shift, rights-based frameworks risk remaining formally robust yet practically inoperative, particularly in low-resource and uneven infrastructural contexts.

\section{Limitations and Conclusion}

While this study undertakes a close comparative analysis of data protection laws and institutional architectures in Bangladesh, our understanding is necessarily shaped by the analytic lenses focused on formal legal texts, regulatory design, and documented enforcement structures, rather than on sustained ethnographic engagement with data subjects or frontline administrators. Although members of the research team bring long-standing familiarity with the region and its political–legal contexts, our engagement with how these laws are experienced in everyday life is mediated through secondary sources, public records, and institutional documentation. As a result, this paper approaches data protection regimes from a governance and design perspective, which may not fully capture the lived strategies through which individuals navigate, resist, or reinterpret these systems.
We also acknowledge that our analysis is constrained by the political sensitivity of data governance in the region. State surveillance, national security exemptions, and executive control over regulatory institutions limit the availability of granular, publicly accessible information about enforcement practices and decision-making processes. In several cases, the absence of transparent reporting or accessible case law restricted how deeply we could trace the operation of remedies or oversight in practice. These constraints are not merely methodological limitations, but themselves reflect the governance conditions the paper seeks to analyze.
Finally, this study deliberately situates its claims within a bounded regional and temporal context. By using Bangladesh as an analytic frame, the paper prioritizes structural comparison over broad generalization. While similar dynamics may be present in other socioeconomic settings, we refrain from extending our claims beyond the specific legal regimes and institutional conditions examined here. Our argument is not that data protection universally fails, but that it fails in patterned ways under particular configurations of power, capacity, and governance design.
Despite these limitations, we believe this study offers a meaningful contribution to HCI and ICTD scholarship by making visible how rights-based data protection can operate as a symbolic and distributive governance technology rather than a mechanism of constraint. By foregrounding mediation, infrastructural limits, and state power, the paper moves beyond implementation-centric critiques and instead reframes data protection as a sociotechnical design problem. In doing so, it provides a foundation for future empirical and design-oriented work that takes these structural conditions seriously when imagining more accountable forms of data governance.

\bibliographystyle{ACM-Reference-Format}
\bibliography{0.main}

\end{document}